\documentclass[%
aps,
 reprint,
 amsmath,amssymb,
prper,
floatfix
]{revtex4-2}

\usepackage{graphicx}
\usepackage{dcolumn}
\usepackage{bm}

\begin{document}


\title{Can an AI-tool grade assignments in an introductory physics course?}

\author{Gerd Kortemeyer}
 \email{kgerd@ethz.ch}
 \affiliation{%
 Educational Development and Technology, ETH Zurich, Zurich, Switzerland
}%
\altaffiliation[also at ]{Michigan State University, East Lansing, USA}

\date{\today}

\begin{abstract}
Problem solving is an integral part of any physics curriculum, and most physics instructors would likely agree that the associated learner competencies are best assessed by considering the solution path: not only the final solution matters, but also how the learner arrived there. Unfortunately, providing meaningful feedback on written derivations is much more labor and resource intensive than only grading the outcome: currently, the latter can be done by computer, while the former involves handwritten solutions that need to be graded by humans. This exploratory study proposes an AI-assisted workflow for grading written physics-problem solutions, and it evaluates the viability of the actual grading step using GPT-4. It is found that the AI-tool is capable of providing  feedback that can be helpful in formative assessment scenarios, but that for summative scenarios, particularly those that are high-stakes, it should only be used for an initial round of grading that sorts and flags solution approaches.
\end{abstract}

\maketitle

\section{Introduction}
\subsection{Generative Pre-trained Transformer}
In fall 2022, Generative Pre-trained Transformer (GPT)~\cite{chatgpt} rapidly gained the World stage as a publicly available AI-tool with surprising capabilities. ChatGPT is a text-based interface to this underlying Large-Language Model (LLM). GPT is essentially a tool that produces plausible fiction using a neural-network-based autocomplete algorithm; in that respect, it is similar to the autocomplete on smartphones that suggests likely next words when the user is composing text messages. However, GPT was trained based on a massive text corpus gleaned from public and proprietary sources, and it was extensively fine-tuned  by humans. Also, GPT does not work with words, but with finer-grained tokens, which are similar but not identical to syllables in a word.

ChatGPT allows for text-based dialogues: prompts by the user, responses by GPT. These dialogues are remarkably human-like, and the system would likely pass the Turing test for many situations~\cite{turing1950}. While being used, apart from continued human training by the company behind GPT, OpenAI, the system does not learn anymore as a whole. However, it learns within the confines of a particular dialogue, so it can refer to statements made earlier in the same dialogue. At some point, though, it hits its internal so-called ``token limit'' --- it can only keep a limited number of tokens in memory, comparable to the working memory limitations of a human. Dialogues thus cannot become too extensive before either an error message occurs or the system simply appears to forget what was stated earlier in the dialogue (which will turn out to be a limitation for this study).

The massive training effort behind GPT resulted in the emergence of capabilities that are not necessarily expected from a language model. GPT-4 has been found to pass several standardized exams in the upper percentiles~\cite{kung2022,lawexam,gpt4}. A previous version, GPT-3, could already pass an introductory physics course at a nominal level~\cite{kortemeyer23ai}, and there are indications that GPT-4 performs even better on physics concepts~\cite{west2023ai}. 

While ChatGPT is limited to text-based input and output, GPT-4 itself is multimodal, so it can accept image input~\cite{gpt4}. Among other examples, in advertising videos, the company demos how handwriting and hand-written mathematical formulas are turned into machine-readable documents. Using this feature, however, requires access to its Application Programming Interface (API), which is unfortunately restricted by OpenAI --- as of April 2023, there is a long waiting list, and thus this functionality could not be tested in this study. 

\subsection{Solving Physics Problems}
The idea of grading problem solutions using AI has been around for a while~\cite{mitros2013}, but was oftentimes hampered by the need to train the AI for specific problems; in other words, the AI learned from humans how to grade a specific problem, and in the end it mimics their scoring of that problem. GPT and similar ``pre-trained'' systems promise a more universal solution, which is able to {\it ad hoc} grade problems it has not encountered before. The fulfillment of this promise seems within reach, since these same ``pre-trained'' systems are also able to solve a wide variety of problems they have not encountered before~\cite{kortemeyer23ai}.

Strategically solving physics problems requires logical, conceptual, and mathematical competencies~\cite{reif1976,reif1995}, and hardly a topic in Physics Education Research has been investigated more extensively~\cite{hsu2004}. While the final solution to physics problems, as well as pre-determined, scaffolded steps along the way, can be assessed to varying degrees using computer systems~\cite{kashyd01,kortemeyer08,risley2001,stelzer2001,dufresne02,fredericks2007,richards2011,perdian2013}, a deeper analysis of the complex problem-solving competencies of learners requires an evaluation of the solution path and derivation~\cite{docktor2016,burkholder2020}. However, grading written solutions currently requires human effort, and the effect of this more meaningful feedback on final learning outcomes may be curbed by having fewer such opportunities due to limited human-grading resources and the time delay before such feedback is returned~\cite{bonham2003}. The purpose of this study is to explore if meaningful feedback can be received from written homework and exam problems while in addition having the benefits of more frequent formative assessment with immediate feedback traditionally afforded by solution-focussed online systems.

\section{Scenario}
\begin{figure*}
\begin{center}
\includegraphics[width=0.8\textwidth]{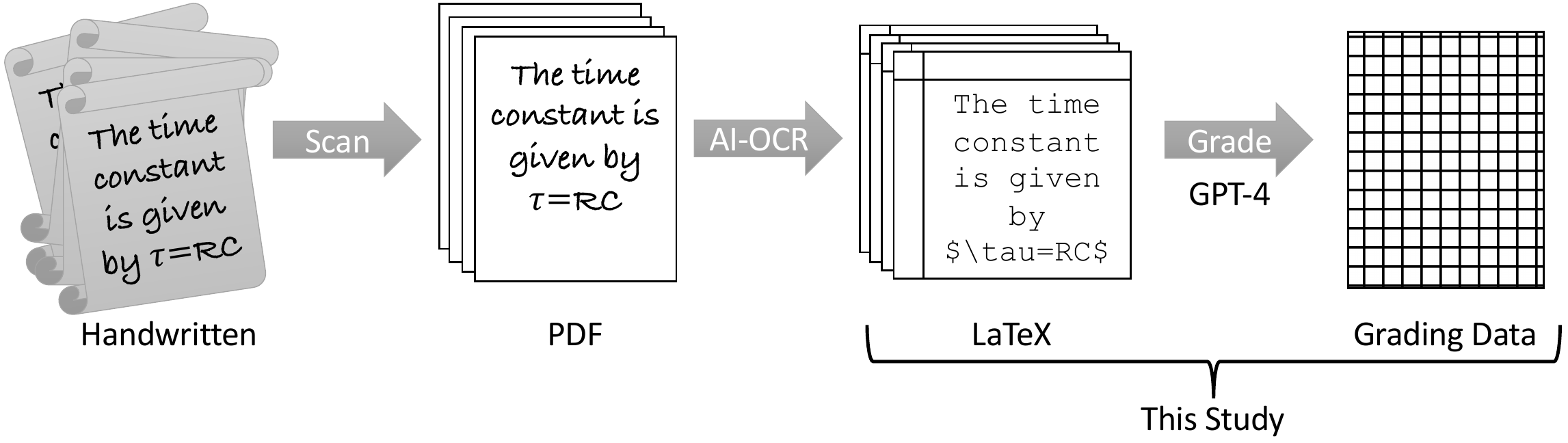}\vspace*{-3mm}
\end{center}
\caption{Possible workflow for AI-assisted grading of handwritten work.\label{fig:workflow}}
\end{figure*}

The proposed scenario has multiple steps, illustrated in Fig.~\ref{fig:workflow}.  In one possible setting, learners are writing problem solutions on paper; while this might seem quaint, most anybody would agree that typing mathematical formulas is cumbersome, and assembling them in formula editors like the ones built into Microsoft Word is time-consuming and tends to be frustrating. The established standard for typing mathematics is LaTeX, and while some experts can ``think'' in LaTeX, paper and pencil are far more efficient and intuitive (there are even arguments that this materiality is essential~\cite{greiffenhagen2014}). This scenario is particular well-suited for exams, which can then take place in traditional, supervised, ``offline'' modes.

\begin{enumerate}
\item The first step is simply scanning the assignments papers into a PDF or image format; this could be accomplished with most copy machines using automatic paper feeding of the student worksheets, where the real challenge are likely paper jams from crumbled up exam sheets.

\item The PDF is then processed by AI-based optical character recognition and translated into a machine-readable format, for example LaTeX. In this study, due to lack of access to the appropriate API, these steps are simulated; the reliability of this step would be the subject of a separate, future study.

\item The LaTeX documents is then graded by an AI-tool, in this study GPT-4. This step might entail several, independent rounds of grading, which are subsequently summarized. Variations include flagging solutions where grading results diverge for subsequent human arbitration, or categorizing the solutions into similarity classes as a preliminary step to more rapid human grading.
\end{enumerate}

A possible second scenario for the first step could be using pen-computing, where learners are directly drawing on a screen. Some modern tablets and laptops are mimicking the paper feel with the appropriate friction, and students appear to be comfortable with this technology, seeing how many of them are using this for lecture notes. A third scenario for the first step would be to photograph or scan the solutions with a smartphone and uploading them to an online system.

These latter two scenarios are less fit for exam settings, as being online means having access to all kinds of resources, communication channels, and online tools (including AI-tools!). These additional affordances of being online would either need to be incorporated into the tasks themselves (possibly making them more demanding) or attempted to be blocked by lock-down technologies~\cite{seb}.

The AI-component of a system like this is very unlikely installable on-premise, and likely only commercial cloud-based solutions are viable. This will raise privacy and data-security concerns that need to be addressed, particularly when dealing with mandatory assignments~\cite{alharthi2015}.

\section{Methodology}
\subsection{Generation of Sample Solutions}
As an example for this study, a simple time-dependent RC-circuit problem was chosen. This problem, shown as the prompt in Fig.~\ref{fig:s1}, involves some conceptual, strategic, and mathematical challenges, but is likely in one form or the other a part of most calculus-based introductory physics courses. The initial potential difference across the capacitor is given, even though it is not needed; this superfluous information makes it tempting to immediately calculate the initial current through the capacitor, leading to a less than straightforward solution, since that current later drops out. Particularly when not working symbolically, but immediately ``plugging and chugging,'' this involves unnecessary steps~\cite{kortemeyer2016bat}.

Since GPT uses a probabilistic algorithm, presenting the same prompt twice will lead to different responses. This property was used to generate a set of 25~unique sample solutions for this study. GPT-4 has better reasoning capabilities than its predecessors, so it more likely produces correct solutions. To also have plausible, but incorrect solutions in the sample, the majority of the solutions were generated by earlier releases. 

At the time of this study (mid-April~2023), three different models of GPT were available through ChatGTP: GPT-4 (March~23, 2023 release~\cite{chatrelease}), Default~GPT-3.5, and Legacy~GPT-3.5. In particular, based on the prompt in Fig.~\ref{fig:s1}. Solutions 1--5 were generated by GPT-4, Solutions 6--13 by Default~GPT-3.5, and Solutions 14--25 by Legacy~GPT-3.5; these solutions are listed in Figs.~\ref{fig:s1}--\ref{fig:s21_25}.

As output format, LaTeX was chosen, since it would be a likely output of the AI-based optical character recognition (see Fig.~\ref{fig:workflow}). During the output phase, ChatGPT converts display equations into a symbol font, so upon copying these expressions into a text file, the LaTeX source code would be lost; thus inline-expressions were requested. Workarounds like these would not be necessary when accessing GPT directly through an API.

\begin{figure}
\begin{center}
\begin{tabular*}{0.46\textwidth}{p{0.46\textwidth}}
\noindent\textbf{Prompt:}
At $t=0$, a resistor (resistance $R$=10 Ohms) is connected across a fully charged capacitor (voltage $V$=5V, capacitance $C$=100mF). At what time $t_2$ does half the initial current flow through the resistor? Do not use display LaTeX, only inline LaTeX.\vspace*{3mm}
\\
\input{solution1}
\end{tabular*}\vspace*{-5mm}
\end{center}
\caption{Prompt and Solution~1 for this study (generated by GPT-4).\label{fig:s1}}
\end{figure}

\begin{figure*}
\begin{center}
\begin{tabular*}{\textwidth}{@{\extracolsep{\fill}}p{0.49\textwidth}  p{0.49\textwidth}}
\input{solution2}
\input{solution3}
&
\input{solution4}
\input{solution5}
\end{tabular*}\vspace*{-6mm}
\end{center}
\caption{Solutions~2 and~5 to the prompt in Fig.~\ref{fig:s1}.\label{fig:s2_5}}
\end{figure*}

\begin{figure*}
\begin{center}
\begin{tabular*}{\textwidth}{@{\extracolsep{\fill}}p{0.49\textwidth}  p{0.49\textwidth}}
\input{solution6}\vspace*{-2mm}
\input{solution7}\vspace*{-2mm}
\input{solution8}
&
\input{solution9}
\input{solution10}
\end{tabular*}\vspace*{-6mm}
\end{center}
\caption{Solutions~6 through~10 to the prompt in Fig.~\ref{fig:s1}.\label{fig:s6_10}}
\end{figure*}

\begin{figure*}
\begin{center}
\begin{tabular*}{\textwidth}{@{\extracolsep{\fill}}p{0.49\textwidth}  p{0.49\textwidth}}

\input{solution11}\vspace*{-2mm}
\input{solution12}
&

\input{solution13}
\input{solution14}
\end{tabular*}\vspace*{-6mm}
\end{center}
\caption{Solutions~11 through~14 to the prompt in Fig.~\ref{fig:s1}.\label{fig:s11_14}}
\end{figure*}

\begin{figure*}
\begin{center}
\begin{tabular*}{\textwidth}{@{\extracolsep{\fill}}p{0.49\textwidth}  p{0.49\textwidth}}
\input{solution15}
\input{solution16}
\input{solution17}
&
\input{solution18}
\input{solution19}
\input{solution20}
\end{tabular*}\vspace*{-6mm}
\end{center}
\caption{Solutions~15 through~20 to the prompt in Fig.~\ref{fig:s1}.\label{fig:s15_20}}
\end{figure*}

\begin{figure*}
\begin{center}
\begin{tabular*}{\textwidth}{@{\extracolsep{\fill}}p{0.49\textwidth}  p{0.49\textwidth}}
\input{solution21}
\input{solution22}
\input{solution23}
&
\input{solution24}
\input{solution25}
\end{tabular*}\vspace*{-6mm}
\end{center}
\caption{Solutions~21 through~25 to the prompt in Fig.~\ref{fig:s1}.\label{fig:s21_25}}
\end{figure*}

Arguably, these samples are setting up GPT to grade itself. While being somewhat verbose and adopting the Royal `We' of instructors and textbooks, for the purposes of this exploratory study, the solutions seem representative enough of what students might submit. This spans the gamut from solutions~1 and~12, which are almost perfect, to solutions~9 and~19, which are completely missing the mark, and it includes the expected unnecessary calculations and transfer of numerical values from one formula to the next. As an aside, while generated to simulate human problem solving, this sample set illustrates the progression in reasoning capabilities between Legacy~GPT-3.5 and GPT-4, achieved in just a few months. These solutions now provide the base for the actual grading study (last step in Fig.~\ref{fig:workflow}).

\subsection{Grading of Solutions}
GPT-4 solved the problem correctly in all instances, so it seems appropriate to use it for grading.
The solutions were scored on a scale of 0~(worst) to 4~(best) on a rubric of correctness of approach, correctness of symbolic derivations, correctness of the numerical result, and straightforwardness. In addition, for each solution, a one-sentence feedback was requested. The rubric scores were combined to a total score with a stronger weight on the final, numerical result. Finally, the system was prompted to generate a correlation table between the solutions based on similarity-of-approach; Fig.~\ref{fig:gradeprompt} shows an example of the associated prompt.

Since GPT is probabilistic, it will not arrive at one deterministic score combination for each solution. Similar to having more than one human grader look at the same solution, each solution was scored several times. Due to the limitations of GPT's token limit, not all 25~solutions could be fed into the system at the same time. They were thus processed in 75~dialogues of randomly compiled batches of 5~solutions, leading to an average of 15~grading rounds for each solution. Unfortunately, more often than not, ChatGPT did not provide the tables in the requested CSV-format, which necessitated some subsequent manual reformatting of the responses in a text editor and Excel. The narrative one-sentence feedback messages from the on the average 15~``graders'' were summarized using GPT-4 into one longer statement reflecting the majority, using the prompt shown in Fig.~\ref{fig:sumprompt}.

\begin{figure}
\begin{center}
\begin{tabular*}{0.46\textwidth}{p{0.46\textwidth}}
\input{gradeprompt}
\end{tabular*}\vspace*{-5mm}
\end{center}
\caption{Example of a grading prompt.\label{fig:gradeprompt}}
\end{figure}

\begin{figure}
\begin{center}
\begin{tabular*}{0.46\textwidth}{p{0.46\textwidth}}
\input{summarizeprompt}
\end{tabular*}\vspace*{-5mm}
\end{center}
\caption{Example of a prompt to summarize the grading feedback.\label{fig:sumprompt}}
\end{figure}

The same solutions were independently graded by the author on the same rubric of correctness of argument, symbolic derivations, and numerical results, as well as straightforwardness. Of course, human grading decisions are also somewhat arbitrary, but this was not further considered in this study. The reader can, however, form his or her own judgement for each solution, which can be found in Figs.~\ref{fig:s1}--\ref{fig:s21_25}.

Several measures of similarity between solutions were considered as basis for clustering: similarity between the solution scores on the rubric for both human and AI grades, as well as the similarity-of-approach provided in the response of the AI-tool; since due to the token limit, only five solutions could be graded at a time, the overall $25\times25$-matrix was compiled by averaging the $5\times5$-matrices. Clustering was accomplished using the gplots-package~\cite{gplots} within R~\cite{rproject}.

\section{Results}

\subsection{Grading Results}
Tables~\ref{tab:ratingtable1_10},~\ref{tab:ratingtable10_20} and~\ref{tab:ratingtable21_25} show the rubric-grading results by GPT-4, based on an average of 15~rounds of grading within different batches of five problems each.

\begin{table*}
\caption{\label{tab:ratingtable1_10}AI-grading of Solutions 1--10.}
\begin{ruledtabular}
\begin{tabular}{lllllp{10.6cm}}
Sol.&Arg.&Symb.&Num.&Str.forw.&Feedback\\\hline
1	&$4.0\pm0.0$	&$4.0\pm0.0$	&$3.9\pm0.3$	&$4.0\pm0.0$	&{\footnotesize The majority of graders agree that Solution 1 is correct, straightforward, and well-structured. The solution uses the correct formula for the current in an RC circuit, derives the expression for the time $t_2$, and solves for the required time when half the initial current flows through the resistor. One grader notes that the final answer should be in milliseconds, not seconds. Overall, the solution is clear, concise, and logically explained.}\\
2	&$3.6\pm0.9$	&$3.7\pm0.6$	&$2.8\pm1.6$	&$3.8\pm0.4$	&{\footnotesize The majority of graders agree that the solution to problem 2 has a correct and clear approach, using the appropriate RC circuit equations and argumentation. However, there seems to be some disagreement on the numerical calculation for $t_2$. The majority opinion indicates that there is an error in the calculation of the time constant, leading to an incorrect numerical result for $t_2$. Overall, the solution is well-organized and logically explained, but the numerical value should be revised.}\\
3	&$3.7\pm0.5$	&$3.7\pm0.5$	&$2.2\pm1.6$	&$3.7\pm0.5$	&{\footnotesize The majority of graders agree that Solution 3 has a correct and straightforward approach, with correct argumentation and symbolic representation. However, there is a common issue with the numerical calculation for $t_2$, with some graders indicating the result is off by a factor of 1000 or in the wrong unit (milliseconds instead of seconds). Despite these errors, the solution is generally well-explained and follows a logical approach.}\\
4	&$3.7\pm0.5$	&$3.8\pm0.4$	&$3.3\pm1.2$	&$3.8\pm0.4$	&{\footnotesize Solution 4 demonstrates a correct and straightforward approach to solving problem 4, using appropriate RC circuit equations and accurately deriving the time $t_2$. However, the majority of the graders point out a minor numerical error in the final answer, possibly due to rounding or unit conversion. Despite this, the solution is clear, well-argued, and symbolically correct.}\\
5	&$3.9\pm0.4$	&$3.9\pm0.4$	&$3.9\pm0.4$	&$3.9\pm0.4$	&{\footnotesize Solution 5 is generally correct and straightforward, successfully applying the exponential decay equation for the RC circuit to derive the result for $t_2$. The majority of the graders appreciate the clear argumentation, symbolic correctness, and numerical accuracy. However, there is a minor numerical error due to rounding, and some graders suggest that the clarity and straightforwardness of the argument could be improved. Overall, the solution demonstrates a good understanding of Ohm's law, the time constant, and the current equation for an RC circuit.}\\
6	&$3.8\pm0.4$	&$3.8\pm0.4$	&$3.3\pm1.1$	&$3.8\pm0.4$	&{\footnotesize The majority of the graders agree that the solution to problem 6 is correct, well-argued, and straightforward in its approach, with accurate symbolic representation and explanation. However, some graders noted an incorrect numerical value in the final result or a mistake in calculating $t_2$.}\\
7	&$3.1\pm0.9$	&$3.6\pm0.6$	&$3.1\pm1.1$	&$3.1\pm0.9$	&{\footnotesize The majority of the graders found Solution 7 to be correct with accurate argumentation, symbolic representation, and numerical values. However, some graders mentioned issues such as incorrect current direction, voltage calculation, and final result, as well as an incorrect manipulation of the equation. Despite these discrepancies, most graders agreed that the solution was straightforward and used the correct approach.}\\
8	&$3.8\pm0.4$	&$3.8\pm0.4$	&$3.4\pm1.1$	&$3.8\pm0.4$	&{\footnotesize The majority of graders agree that Solution 8 has correct argumentation, symbolic representation, and a straightforward explanation. The solution correctly derives the formula for the time at which half the initial current flows and uses the appropriate approach. However, there is disagreement on the numerical value, with some graders noting that it is incorrect or off by a factor due to not considering units. Overall, the solution is mostly correct, but the final numerical answer may need to be revised.}\\
9	&$0.2\pm0.4$	&$1.0\pm0.9$	&$0.0\pm0.0$	&$0.8\pm0.7$	&{\footnotesize Solution 9 demonstrates an incorrect understanding of the initial current, mistakenly assuming it to be zero. This leads to incorrect calculations, argumentation, and the false conclusion that half the initial current never flows through the resistor. The overall approach and reasoning are flawed, resulting in an incorrect solution.}\\
10	&$2.8\pm1.1$	&$3.1\pm0.8$	&$1.3\pm1.3$	&$2.9\pm0.9$	&{\footnotesize The majority of the graders agree that the solution to problem 10 has correct argumentation and symbolic representation, with a clear and straightforward approach. However, there are errors in the numerical calculations, including the time constant and natural logarithm term, leading to an incorrect final result. Some graders also mentioned unnecessary complexity and incorrect derivations.}
\end{tabular}
\end{ruledtabular}
\end{table*}

\begin{table*}
\caption{\label{tab:ratingtable10_20}AI-grading of Solutions 11--20.}
\begin{ruledtabular}
\begin{tabular}{lllllp{10.6cm}}
Sol.&Arg.&Symb.&Num.&Str.forw.&Feedback\\\hline
11	&$3.5\pm0.8$	&$3.5\pm0.7$	&$2.8\pm1.4$	&$3.5\pm0.8$	&{\footnotesize The majority of the graders agree that Solution 11 has a correct approach by using the exponential decay equation for the RC circuit and deriving the equation for i(t). However, there seems to be a consensus that there are errors in the calculations, leading to an incorrect numerical result for $t_2$. The argumentation, symbolic correctness, and straightforwardness could be improved. Despite these errors, some graders still consider the solution to be well-structured and clear.}\\
12	&$4.0\pm0.0$	&$4.0\pm0.0$	&$4.0\pm0.0$	&$4.0\pm0.0$	&{\footnotesize Solution 12 is a correct, well-structured, and straightforward approach to the problem. The majority of graders praised the clear argumentation, appropriate use of symbols and equations, and accurate numerical results. The solution effectively uses the RC circuit formula and provides a concise explanation.}\\
13	&$3.9\pm0.3$	&$3.9\pm0.3$	&$3.5\pm1.0$	&$3.9\pm0.3$	&{\footnotesize The majority of the graders agree that Solution 13 is correct, clear, and straightforward. The solution uses the appropriate RC circuit equations, symbols, and approach to find the time when half the initial current flows through the resistor. While there is mention of a sign error and incorrect final result by a couple of graders, the overall consensus supports the solution's correctness and organization.}\\
14	&$3.2\pm0.9$	&$3.4\pm0.8$	&$2.3\pm1.4$	&$3.2\pm0.9$	&{\footnotesize The majority of graders agree that Solution 14 has a correct approach and argumentation, with symbolic correctness and straightforwardness. However, there are inconsistencies in the numerical results, likely due to errors in calculations or sign errors.}\\
15	&$2.4\pm1.4$	&$2.4\pm1.4$	&$1.9\pm1.6$	&$2.3\pm1.5$	&{\footnotesize The majority of graders agree that the solution to problem 15 is mostly accurate in terms of approach and equations used, but there are some minor mistakes in argumentation, symbolic correctness, and numerical calculations. The solution is considered to be somewhat straightforward and well-explained, but with a few errors in derivation and numerical results.}\\
16	&$3.3\pm0.8$	&$3.6\pm0.5$	&$1.9\pm1.5$	&$3.6\pm0.5$	&{\footnotesize The majority of the graders agree that Solution 16 has a correct approach and demonstrates accurate symbolic representation, argumentation, and use of formulas. However, a calculation mistake in determining the initial current and subsequent errors in numerical calculations lead to an incorrect numerical answer for $t_2$. Despite these numerical issues, the solution is well-organized and clear.}\\
17	&$3.3\pm1.0$	&$3.8\pm0.6$	&$2.2\pm1.4$	&$3.6\pm0.6$	&{\footnotesize Solution 17 demonstrates the correct approach, argumentation, and symbolic representation, but makes an error in the numerical calculation, specifically a sign error in the time constant, which leads to an incorrect final result. Despite this, the solution is considered straightforward and directly addresses the problem.}\\
18	&$3.1\pm0.8$	&$3.4\pm0.6$	&$2.6\pm1.2$	&$3.2\pm0.7$	&{\footnotesize Solution 18 has the correct approach and argumentation, but the majority of graders point out an incorrect time constant calculation, which leads to an incorrect final numerical result. Despite this, the solution is well-structured and uses proper symbols and numerical values.}\\
19	&$0.3\pm0.6$	&$0.9\pm0.9$	&$0.2\pm0.4$	&$0.5\pm0.7$	&{\footnotesize The majority of the graders agree that Solution 19 contains multiple errors, including incorrect initial current and voltage calculations, as well as an incorrect approach for finding the time $t_2$ when half of the initial current flows through the resistor. These errors lead to an incorrect final result. The solution also has issues with argumentation, symbolic representation, and straightforwardness.}\\
20	&$3.1\pm0.7$	&$3.5\pm0.5$	&$2.0\pm1.3$	&$3.3\pm0.6$	&{\footnotesize Solution 20 generally employs the correct approach and uses the exponential decay equation for solving the problem, but there are errors in the numerical calculations, specifically in the time constant and initial current. These inaccuracies lead to an incorrect final result. While some graders appreciate the clear and straightforward steps, the majority point out the numerical errors affecting the outcome.}
\end{tabular}
\end{ruledtabular}
\end{table*}

\begin{table*}
\caption{\label{tab:ratingtable21_25}AI-grading of Solutions 20--25.}
\begin{ruledtabular}
\begin{tabular}{lllllp{10.6cm}}
Sol.&Arg.&Symb.&Num.&Str.forw.&Feedback\\\hline
21	&$1.6\pm1.1$	&$1.9\pm1.0$	&$0.7\pm0.8$	&$1.8\pm1.1$	&{\footnotesize The majority of the graders agree that Solution 21 has an incorrect approach, incorrect or incomplete argumentation, and incorrect numerical calculations. The relationship between i(t) and dv/dt is also incorrect, leading to a wrong calculation for $t_2$. Some graders acknowledge clear explanation, symbolism, or straightforwardness in parts of the solution, but overall, the solution is deemed incorrect and weak.}\\
22	&$3.7\pm0.6$	&$3.9\pm0.4$	&$3.4\pm1.2$	&$3.7\pm0.6$	&{\footnotesize The majority of graders agree that Solution 22 is correct and straightforward in its approach to finding the time when half the initial current flows through the resistor. The solution uses the time constant and the voltage across the capacitor as a function of time. However, a few graders noted some minor issues with the numerical calculation and representation of the time constant. Overall, the solution is well-explained and clear, with accurate derivation and symbolic correctness.}\\
23	&$2.1\pm1.5$	&$2.6\pm1.1$	&$1.0\pm1.6$	&$2.1\pm1.4$	&{\footnotesize The majority of the graders find Solution 23 to have a correct approach and clear step-by-step argumentation. However, there are concerns about the incorrect final numerical result and potential errors in the calculations involving the exponential term and natural logarithm. Some graders also mention an incorrect use of the time constant, incorrect equation, and wrong numerical value.}\\
24	&$2.9\pm1.2$	&$3.5\pm0.7$	&$1.7\pm1.4$	&$3.0\pm0.9$	&{\footnotesize The majority of graders agree that Solution 24 uses the correct approach, equations, and symbolic representation to find the time at which half the initial current flows through the resistor. However, there are numerical errors in the calculation, particularly regarding the time constant and the voltage across the capacitor. Additionally, some graders noted unnecessary complexity in the solution. Overall, the solution is partially correct but suffers from incorrect numerical values and minor inconsistencies.}\\
25	&$1.7\pm1.3$	&$2.7\pm0.8$	&$0.7\pm1.1$	&$2.1\pm1.1$	&{\footnotesize The majority of the graders agree that Solution 25 has an incorrect numerical result for $t_2$ due to inconsistencies, errors, and incorrect values in the derivation and equation. The approach is mostly correct and some steps are symbolically accurate, but there are sign errors and the method is not straightforward.}
\end{tabular}
\end{ruledtabular}
\end{table*}

Striking are the scores that have more than one point of standard deviations to their averages, which are:
\begin{itemize}
\item The correctness of the argument in Solutions~10,~15,~21,~23,~24, and~25: All of these solutions in the end arrive at the wrong numerical answer. Solutions~10 and~15 also needlessly calculate a numerical value for the initial current, while Solutions~21,~23, and~25 take a slight detour via the capacitative current $I(t)=C\cdot dV/dt$. Solution~24 starts with the expression for a charging instead of a discharging capacitor, which would still have arrived at the same answer had it not been for numerical errors. 
\item The correctness of the symbolic calculations in Solutions~15 and~23: Solution~15 is plug-and-chug, i.e., it plugs numbers into the first available formula and then transfers the numerical result to the next formula. Thus, the score for symbolic correctness of Solution~15 should have been consistently low for all grading rounds. Solution~23 makes an error in the very first line of the symbolic calculations by not eliminating the capacitance $C$.
\item The correctness of the numerical results, in particular for Solutions~2,~3,~4,~6,~7, etc. --- 18~out of the 25~problems altogether: 8~of these solutions actually have the correct numerical answer, 4~are wrong by one or more orders of magnitude, and 6~of the problems have a completely different numerical answers due to various reasons. The large list may be the result of GPT~4 still falling short on numerical calculations.
\item The straightforwardness of Solutions~15,~21,~23, and~25: All of these solutions already appeared in other lists of ambiguous scores. 
\end{itemize}
Overall, there does not appear to be a discernible pattern as to why these solutions are leading to widely spread scores. The one-sentence feedback given by GPT-4 also does not provide helpful hints for determining which features of these solutions the algorithm may have latched on to.

The summaries of the one-sentence feedback message sound very plausible, but many of them include incorrect or at least misleading statements. Examples of incorrect statements include:
\begin{itemize}
\item ``The majority opinion indicates that there is an error in the calculation of the time constant, leading to an incorrect numerical result for $t_2$'' for Solution~2, when in fact the numerical result is correct.
\item ``The majority of graders agree that Solution~24 uses the correct approach, equations, and symbolic representation,'' when in fact the equation for a charging capacitor is used. 
\end{itemize}
Examples of misleading statements are:
\begin{itemize}
\item ``One grader notes that the final answer should be in milliseconds, not seconds'' for Solution~1; while it is true that microseconds would have been more elegant, this is not required.
\item ``Solution 17 demonstrates the correct approach,'' when it fact it includes unnecessary calculations.
\end{itemize}
Often this includes remarks about numerical inaccuracies, even if the result is correct. However, particularly for the solution attempts that were completely incorrect, the feedback can be useful, for example, ``Solution~9 demonstrates an incorrect understanding of the initial current, mistakenly assuming it to be zero.''

\subsection{Agreement with Manual Grading}
\begin{figure}
\begin{center}
\includegraphics[width=\columnwidth]{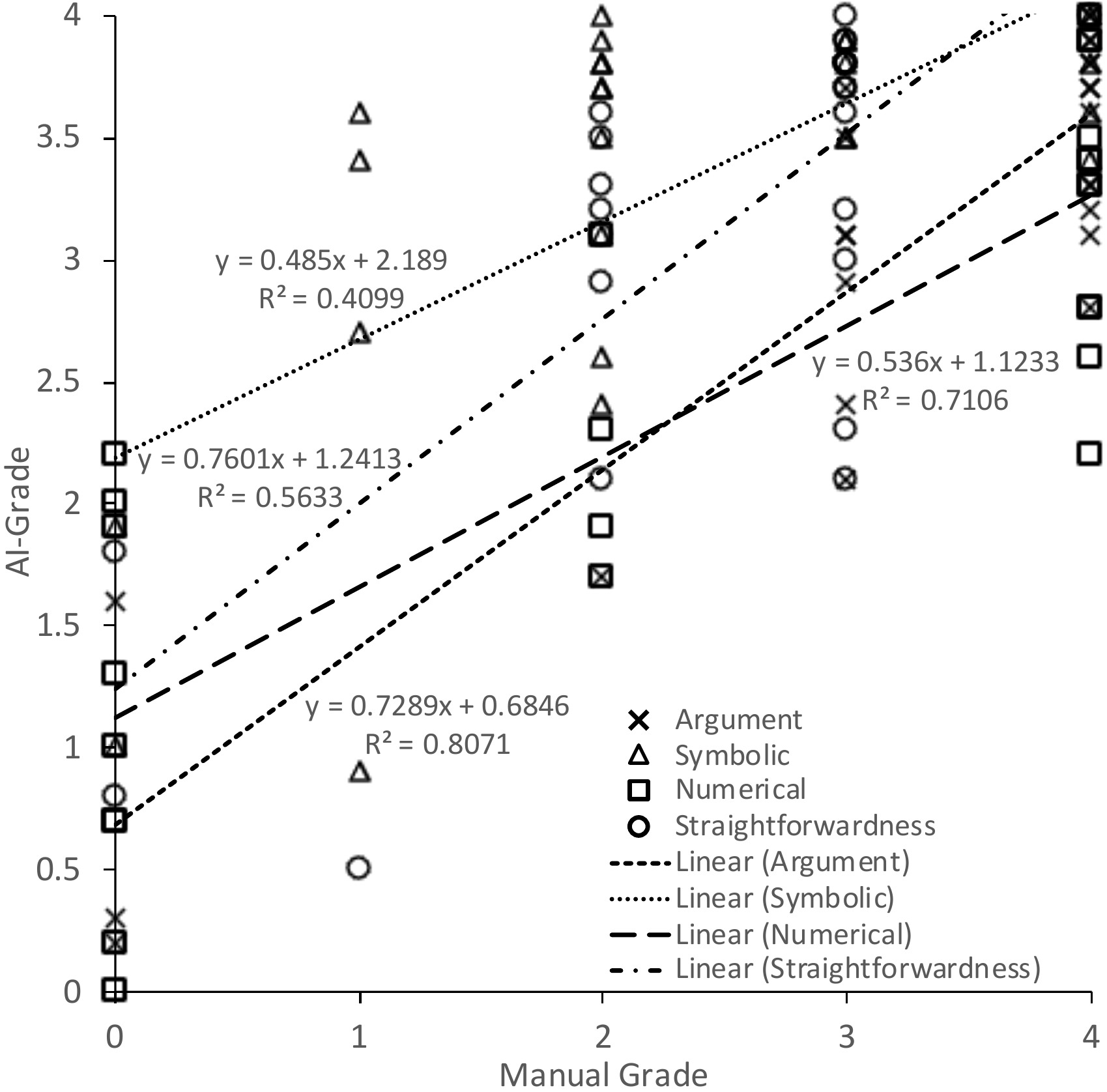}
\end{center}
\caption{Correlation between manual and AI-grading rubric scores.\label{fig:mancor}}
\end{figure}

Fig.~\ref{fig:mancor} shows the correlation between the rubric scores resulting from manual and AI-grading. The scores are clearly positively correlated, with a particularly high $R^2$ for the correctness of the argument and the numerical answer. This finding is surprising, since the correctness of the argument is more subjective than for example the correctness of the symbolic operations, and the scoring of the correctness of the numerical answers is the one with the highest standard deviation between grading cycles. Symbolic correctness has the lowest $R^2$ and is generally rated a lot higher by the AI than by the author.

\begin{figure}
\begin{center}
\includegraphics[width=\columnwidth]{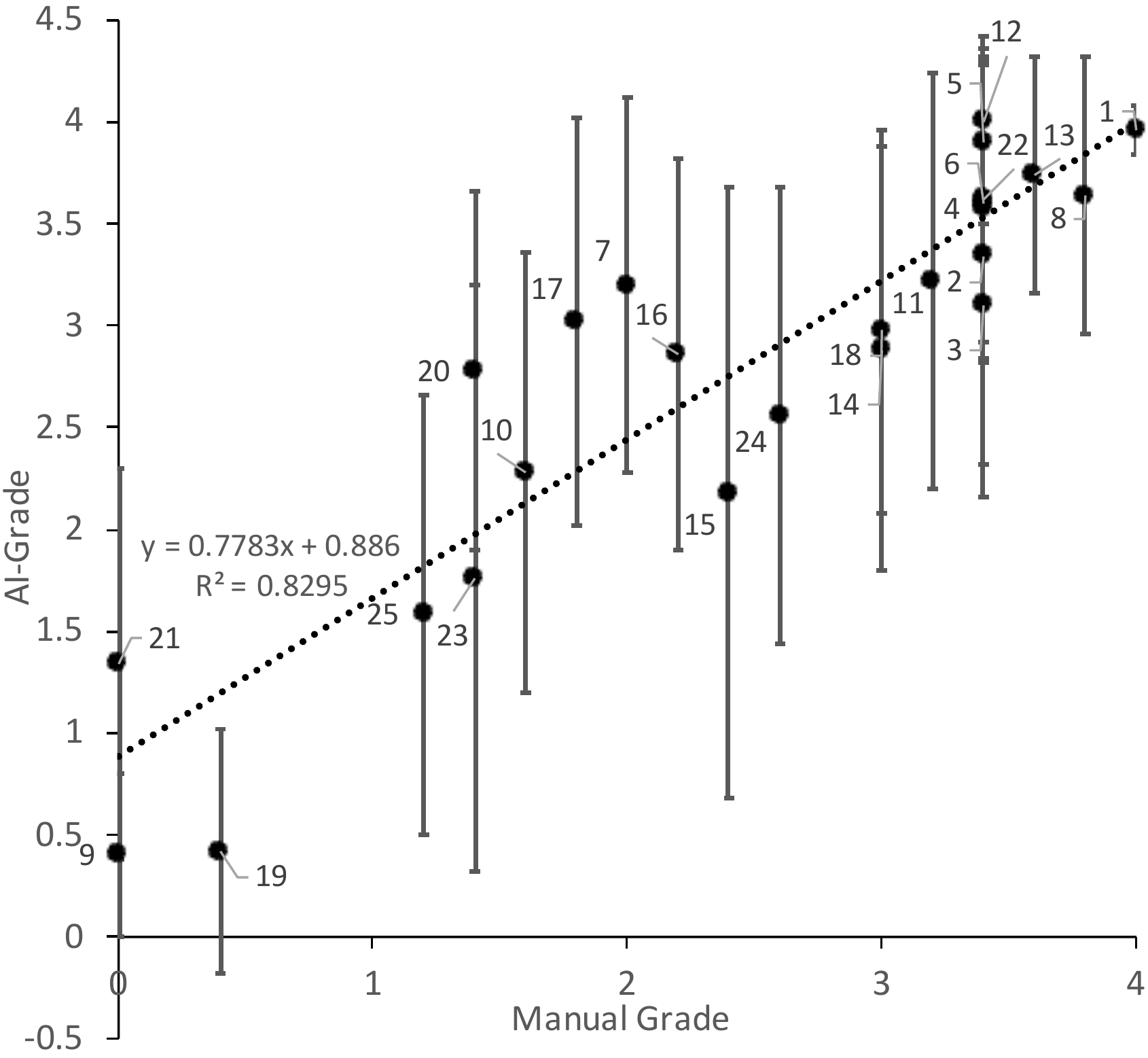}
\end{center}
\caption{Correlation between manual and AI-grading total scores. For each data point, the solution number and combined AI-grading standard deviations are indicated.\label{fig:mantot}}
\end{figure}

Fig.~\ref{fig:mantot} shows the total scores on the problem, where the rubric items approach, symbolic correctness, and straightforwardness are weighted 20\%, and the correct final numerical answer 40\%, respectively. In addition, the correspondingly combined standard deviations for the AI-gradings are given by error bars, and the data points are labeled by solution number.

The scores for the best solutions agree fairly well with the linear interpolation ($R^2\approx0.83$), but fluctuations are higher in partial credit situations.  On the average, GPT assigns total scores that are almost 0.9 points higher than those by the author

\subsection{Clustering}
\begin{figure}
\begin{center}
\includegraphics[width=\columnwidth]{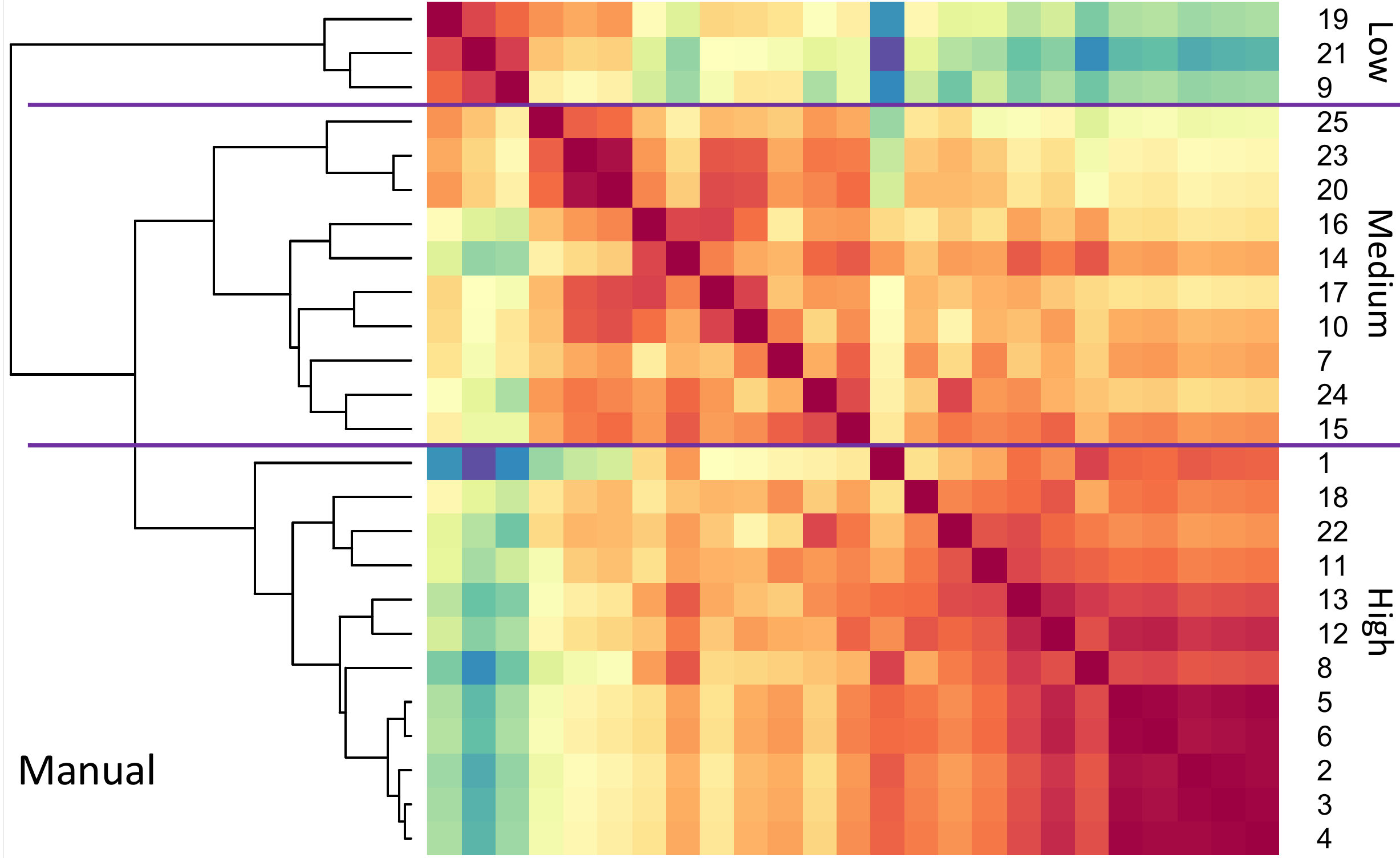}

\includegraphics[width=\columnwidth]{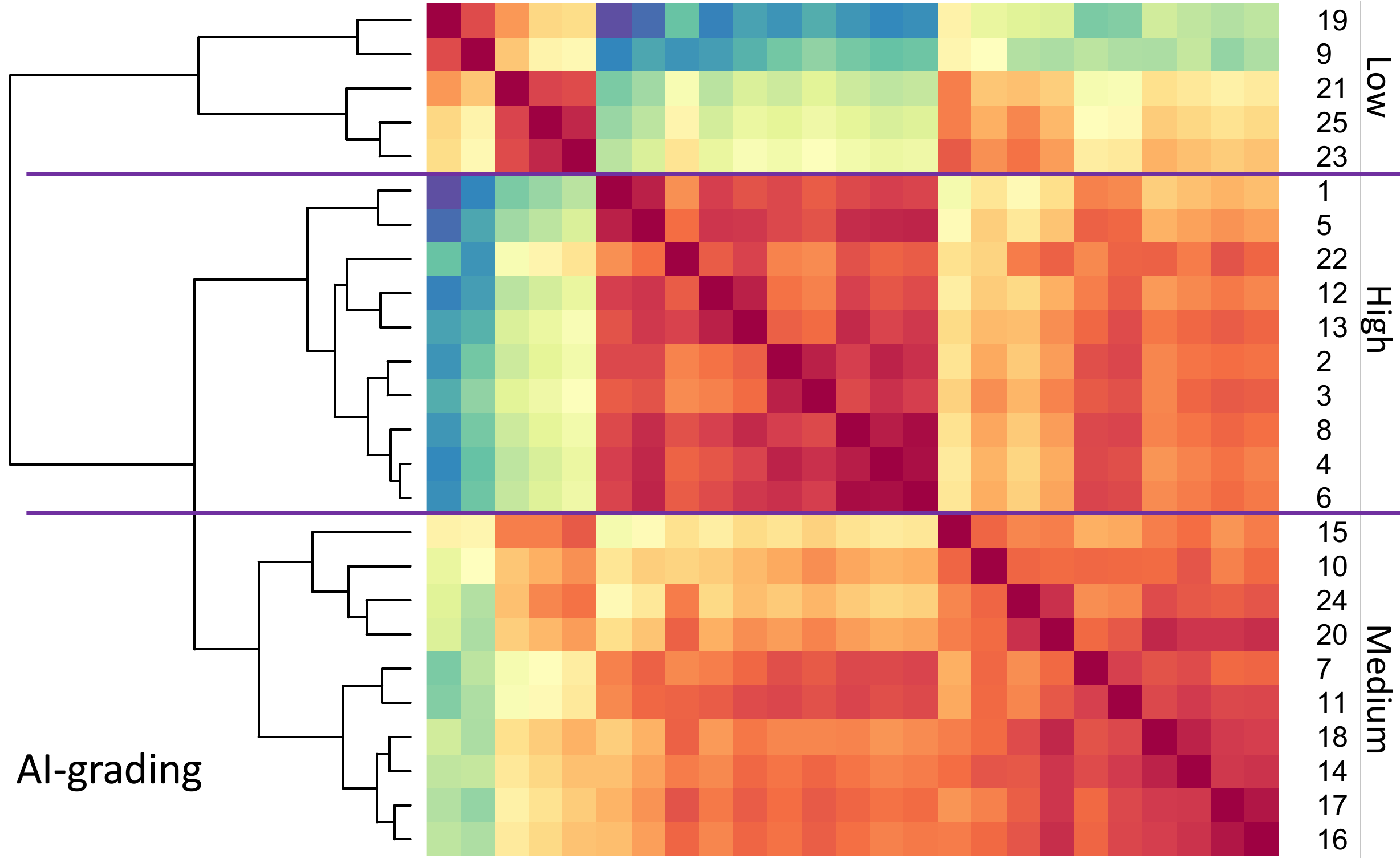}

\includegraphics[width=\columnwidth]{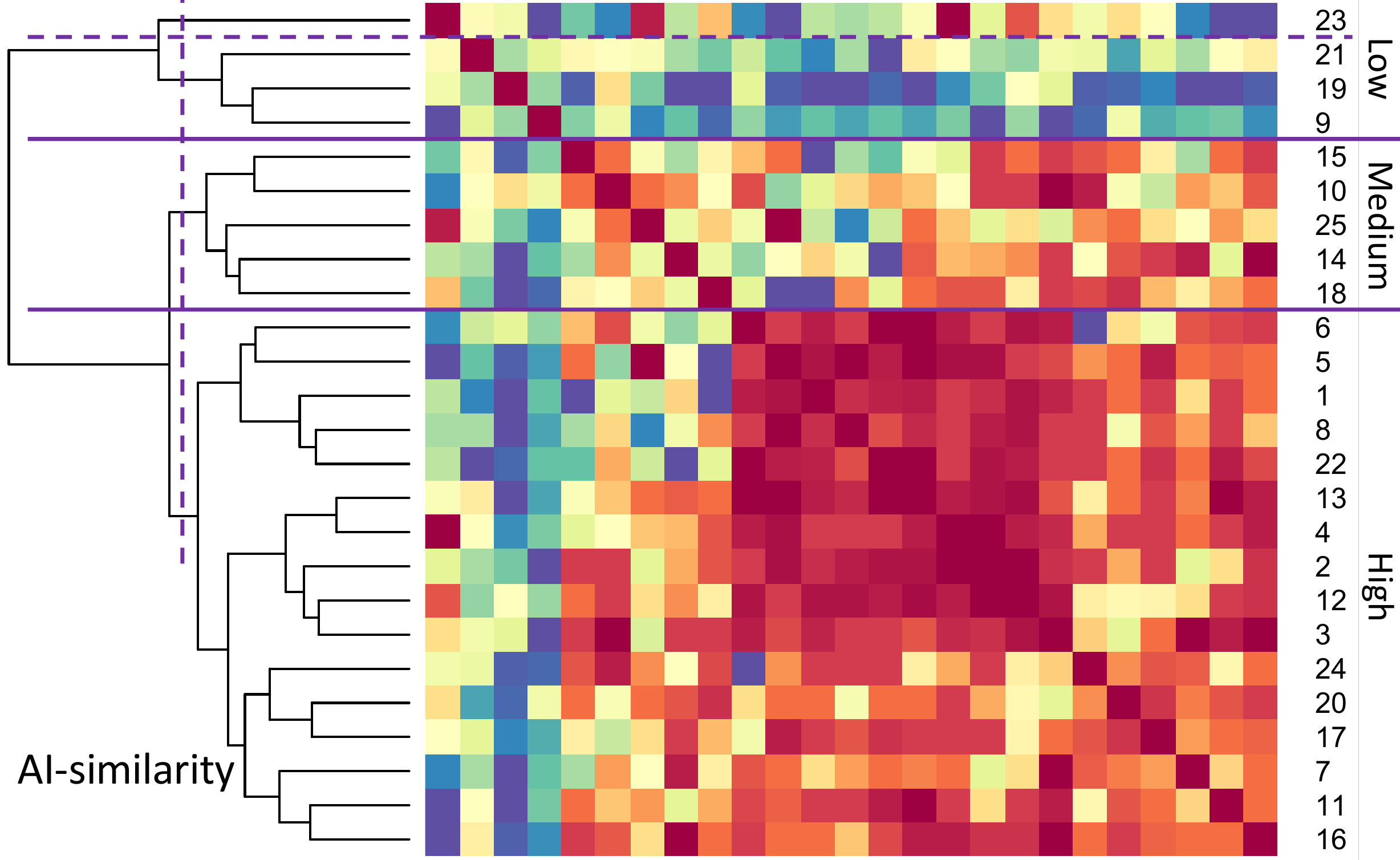}

\end{center}
\caption{Dendrograms and heat maps of similarities between solutions based on manual grading (top panel), AI-grading (middle panel), and AI-approach similarity (bottom panel). The cluster cuts are indicated by purples lines; the dashed line indicates possible cuts.\label{fig:sims}}
\end{figure}

Figure~\ref{fig:sims} shows dendrograms and heat maps of the similarities between the solutions based on manual grading, AI-grading, and AI-approach similarity, respectively. Based on these, three clusters emerge for the manual and the AI-grading. For the AI-similarity-of-approach measure (determined by GPT in response to the prompt in Fig.~\ref{fig:gradeprompt}), either two clusters could be identified, or the dendrogram could be cut at a deeper level (indicated by dashed lines in Fig.~\ref{fig:sims}), resulting in four clusters. However, the fourth cluster only contains Solution~23, so a decision to not treat it separately could be justified.

These clusters are shown in Table.~\ref{tab:clustering}; it turns out that they roughly align with the total points in Fig.~\ref{fig:mantot}, with the low, medium, and highly scored problems forming the clusters. 

\begin{table*}
\caption{\label{tab:clustering}Clustering based on the dendrograms and heat maps in Fig.~\ref{fig:sims}.}
\begin{ruledtabular}
\begin{tabular}{lrcl}
			&Low		&Medium	&High\\\hline
Manual		&9 19 21		&7 10 14 15 16 17 20 23 24 25	&1 2 3 4 5 6 8 11 12 13 18 22\\
AI-grading		&9 19 21 23 25	&7 10 11 14 15 16 17 18 20 24	&1 2 3 4 5 6 8 12 13 22\\
AI-similarity	&9 19 21 (23)	&10 14 15	18 25&1 2 3 4 5 6 7 8  11 12 13 16 17 20 22 24
\end{tabular}
\end{ruledtabular}
\end{table*}

All similarity measures identify the almost entirely wrong solutions~9,~19, and~21 as members of the ``low'' cluster. The manual grading has Solution~23 in the middle cluster, as the only mistake  is $\frac{CV_0}{-RC}=\frac{V_0}{RC}$, that is, an error in the calculation rather than a fundamental error; this solution was singled out in the clustering according to similarity-of-approach. Solution~25 not once but twice includes the same kind of error in calculating fractions.

Solutions~11 and~18 made it into the highest cluster for manual grading, but are not found in the corresponding cluster for AI-grading. Both solutions arrive at the correct result (even though Solution~18 is somewhat nonchalant with the units in the second-to-last expression).

Overall, clustering the solutions based on the similarity of the scores within the grading rubric provided similar information for manual and AI-grading, while clustering based on the similarity-of-approach is less informative. The automatically formed clusters based on AI could be used for sorting solutions in preparation for human grading.

\section{Discussion}
\subsection{Potential for Assisting in Grading}
GPT shows great promise as a tool for grading written solutions to physics problems, as evidenced by the results in Fig.~\ref{fig:mantot}. The system performs best at the extreme ends of the grading spectrum: clearly correct and clearly incorrect solutions are generally reliably recognized, but overall, the system currently falls short of the stringency required for standalone, unsupervised exam grading. Also, the system generally assigns more points than the author would.

While there are proposals to mitigate these shortcomings by providing ``high transparency''  assessments with multiple attempts and regrade requests~\cite{azad2020}, students will likely only invoke these for false negatives, i.e., scores that are lower than what they expected. This could lead to grade inflation. Instead of waiting for student complaints, instructional staff will need to verify the scores before the exam is returned.

As typical for the tool, feedback sounds plausible, but might be misleading. In grading problems, GPT-4 still exhibits some of the same weaknesses with symbolic and numerical calculations that its predecessor GPT-3 did in solving problems~\cite{kortemeyer23ai}. OpenAI promises the optional incorporation of Wolfram Alpha~\cite{walpha}, which may solve this issue, but just like for the API, there is also a waiting list for the required plugin.

\subsection{Potential for Assisting in Learning}
If not used for grading exams, especially in connection with smartphone-based scanning, GPT or similar tools could be used for assisting students in learning. This is a far more classic domain for AI in physics education~\cite{graesser2001,vanlehn2005}

In that context it is important if the system could pinpoint particular errors. At the moment, though, this is still somewhat doubtful. For example, when prompted to find the mistake in Solution~14, the system responds:
\begin{quote}
The mistake in the solution is in the definition of the time constant. The correct definition is $RC = R \cdot C$. The rest of the solution is correct. We need to correct the time constant and then find the correct value of $t_2$.

The time constant is defined as $RC = R \cdot C$. In this case, $R = 10,\text{Ohms}$ and $C = 100,\text{mF} = 100 \cdot 10^{-3},\text{F}$, so the time constant is:

$RC = R \cdot C = 10,\text{Ohms} \cdot 100 \cdot 10^{-3},\text{F} = 1,\text{s}$

Now we can substitute the corrected time constant into the equation for $t_2$:

$t_2 = -RC \ln{\frac{1}{2}} = -1,\text{s} \ln{\frac{1}{2}} \approx 0.693,\text{s}$

So the correct time $t_2$ at which half the initial current flows through the resistor is approximately 0.693 seconds.
\end{quote}
The error is not really in the ``definition'' of the time constant, but simply in the calculation of the powers of ten.

When prompted to pinpoint the error in Solution~23, the response is:
\begin{quote}
The mistake in the solution is in the derivation of the equation $i(t_2) = \frac{1}{2} i(0)$. The correct equation should be:

$i(t_2) = \frac{1}{2} i(0) = \frac{1}{2} \frac{V_0}{R}$

However, the solution incorrectly equates this expression to the equation for $i(t)$:

$-\frac{V_0}{2RC} e^{-\frac{t_2}{RC}}$

The correct equation for $i(t)$ should be:

$i(t) = \frac{V_0}{R} e^{-\frac{t}{RC}}$

Now, we can set up the correct equation for $i(t_2)$:

$\frac{1}{2} i(0) = \frac{1}{2} \frac{V_0}{R} = \frac{V_0}{R} e^{-\frac{t_2}{RC}}$

Now, we solve for $t_2$:

$e^{-\frac{t_2}{RC}} = \frac{1}{2}$

$-\frac{t_2}{RC} = \ln{\left(\frac{1}{2}\right)}$

$t_2 = RC \ln{2} = 10 \text{s} \ln{2} \approx 6.93\text{s}$

So the time $t_2$ at which half the initial current flows through the resistor is approximately 6.93 seconds.
\end{quote}
The system correctly found that the $C$ was not cancelled, but then actually proceeded to wrongly calculate the answer. The feedback given by the system can be helpful, but needs to be evaluated critically by the learner --- which in itself is an increasingly valuable competency, as AI will almost invariably become ubiquitous in everyday life.

In any case, the system should not attempt to give away the solution; unfortunately, using the prompt ``Act as a socratic teacher and assist a learner with this solution attempt'' made the reply more personable and supportive (``You're on the right track, but there is a small mistake \ldots''), but did not stop the system from providing a solution, either.

\section{Limitations}
This study is limited to one particular tool, GPT-4, at one particular point in time, mid-April 2023. Due to the rapid development of AI-tools, it can thus provide nothing more than a snapshot of what is minimally possible. The initial step, conversion of handwritten solutions involving mathematical expressions to a machine-readable format could not be investigated here, due to lack of access to the required Application Programming Interface.

The author had to make choices when providing prompts to GPT. While some experimentation went into formulating the prompts, due to the probabilistic nature of the responses, some decisions were made based on anecdotal evidence, and better prompts framing the solutions could well have resulted in more reliable results. 

GPT is only one of the rapidly evolving tools becoming broadly available. There are competing solutions by Google~\cite{bard}, NVIDIA~\cite{nvidia}, and Microsoft~\cite{microsoft}, which may perform better or worse, but could not be evaluated here.

\section{Outlook}
The next step to this exploratory study would be enabled by access to the API (either of GPT or of other tools on the market) and consist of fully implementing the workflow in Fig.~\ref{fig:workflow}; performance of the optical character recognition is also crucial to the reliability of future systems.

A future study should involve authentic student work instead of GPT grading itself, for example from exams in large-enrollment physics courses, and compare the results from AI-grading to those from a traditional grader pool. Particularly in high-stake exams, grading usually involves more than one round, so inter-rater reliability could also be established for the human graders. As a welcome side-effect, earlier steps in Fig.~\ref{fig:workflow} could provide better workflow management even for human graders, as less paper would be shuffled around and grades could more easily be adjusted.

\section{Conclusion}
In this exploratory  study, GPT has shown considerable potential for grading freeform student work in physics. While AI-assigned grades have a strong correlation to manually assigned grades, they are currently not reliable enough for summative assessments, such as high-stake exams. The system, however, is reliable enough to assist human graders by pre-sorting or clustering solutions and by providing preliminary scores. GPT still remains hampered by its limited capabilities and inconsistencies carrying out symbolic and numerical calculations, so currently several independent grading rounds are needed. The narrative feedback provided by the system seems plausible, but currently still frequently falls short of being reliable. The system can be helpful in formative assessment, but also in that scenario, learners need to critically evaluate its responses.

\begin{acknowledgments}
The author would like to thank Christine Kortemeyer for helpful feedback.
\end{acknowledgments}

\bibliography{ChatGPTGrade}

\begin{thebibliography}{36}%
\makeatletter
\providecommand \@ifxundefined [1]{%
 \@ifx{#1\undefined}
}%
\providecommand \@ifnum [1]{%
 \ifnum #1\expandafter \@firstoftwo
 \else \expandafter \@secondoftwo
 \fi
}%
\providecommand \@ifx [1]{%
 \ifx #1\expandafter \@firstoftwo
 \else \expandafter \@secondoftwo
 \fi
}%
\providecommand \natexlab [1]{#1}%
\providecommand \enquote  [1]{``#1''}%
\providecommand \bibnamefont  [1]{#1}%
\providecommand \bibfnamefont [1]{#1}%
\providecommand \citenamefont [1]{#1}%
\providecommand \href@noop [0]{\@secondoftwo}%
\providecommand \href [0]{\begingroup \@sanitize@url \@href}%
\providecommand \@href[1]{\@@startlink{#1}\@@href}%
\providecommand \@@href[1]{\endgroup#1\@@endlink}%
\providecommand \@sanitize@url [0]{\catcode `\\12\catcode `\$12\catcode
  `\&12\catcode `\#12\catcode `\^12\catcode `\_12\catcode `\%12\relax}%
\providecommand \@@startlink[1]{}%
\providecommand \@@endlink[0]{}%
\providecommand \url  [0]{\begingroup\@sanitize@url \@url }%
\providecommand \@url [1]{\endgroup\@href {#1}{\urlprefix }}%
\providecommand \urlprefix  [0]{URL }%
\providecommand \Eprint [0]{\href }%
\providecommand \doibase [0]{https://doi.org/}%
\providecommand \selectlanguage [0]{\@gobble}%
\providecommand \bibinfo  [0]{\@secondoftwo}%
\providecommand \bibfield  [0]{\@secondoftwo}%
\providecommand \translation [1]{[#1]}%
\providecommand \BibitemOpen [0]{}%
\providecommand \bibitemStop [0]{}%
\providecommand \bibitemNoStop [0]{.\EOS\space}%
\providecommand \EOS [0]{\spacefactor3000\relax}%
\providecommand \BibitemShut  [1]{\csname bibitem#1\endcsname}%
\let\auto@bib@innerbib\@empty
\bibitem [{\citenamefont {{OpenAI}}(2023{\natexlab{a}})}]{chatgpt}%
  \BibitemOpen
  \bibfield  {author} {\bibinfo {author} {\bibnamefont {{OpenAI}}},\
  }\href@noop {} {\bibinfo {title} {{ChatGPT}}},\ \bibinfo {howpublished}
  {\url{https://chat.openai.com/chat}} (\bibinfo {year} {accessed April
  2023}{\natexlab{a}})\BibitemShut {NoStop}%
\bibitem [{\citenamefont {Turing}(1950)}]{turing1950}%
  \BibitemOpen
  \bibfield  {author} {\bibinfo {author} {\bibfnamefont {A.~M.}\ \bibnamefont
  {Turing}},\ }\bibfield  {title} {\bibinfo {title} {Computing machinery and
  intelligence},\ }\href {https://doi.org/10.1093/mind/LIX.236.433} {\bibfield
  {journal} {\bibinfo  {journal} {Mind}\ ,\ \bibinfo {pages} {433}} (\bibinfo
  {year} {1950})}\BibitemShut {NoStop}%
\bibitem [{\citenamefont {Kung}\ \emph {et~al.}(2022)\citenamefont {Kung},
  \citenamefont {Cheatham}, \citenamefont {Medinilla}, \citenamefont {ChatGPT},
  \citenamefont {Sillos}, \citenamefont {De~Leon}, \citenamefont {Elepano},
  \citenamefont {Madriaga}, \citenamefont {Aggabao}, \citenamefont
  {Diaz-Candido} \emph {et~al.}}]{kung2022}%
  \BibitemOpen
  \bibfield  {author} {\bibinfo {author} {\bibfnamefont {T.~H.}\ \bibnamefont
  {Kung}}, \bibinfo {author} {\bibfnamefont {M.}~\bibnamefont {Cheatham}},
  \bibinfo {author} {\bibfnamefont {A.}~\bibnamefont {Medinilla}}, \bibinfo
  {author} {\bibnamefont {ChatGPT}}, \bibinfo {author} {\bibfnamefont
  {C.}~\bibnamefont {Sillos}}, \bibinfo {author} {\bibfnamefont
  {L.}~\bibnamefont {De~Leon}}, \bibinfo {author} {\bibfnamefont
  {C.}~\bibnamefont {Elepano}}, \bibinfo {author} {\bibfnamefont
  {M.}~\bibnamefont {Madriaga}}, \bibinfo {author} {\bibfnamefont
  {R.}~\bibnamefont {Aggabao}}, \bibinfo {author} {\bibfnamefont
  {G.}~\bibnamefont {Diaz-Candido}}, \emph {et~al.},\ }\bibfield  {title}
  {\bibinfo {title} {Performance of chatgpt on usmle: Potential for ai-assisted
  medical education using large language models},\ }\href@noop {} {\bibfield
  {journal} {\bibinfo  {journal} {medRxiv}\ ,\ \bibinfo {pages} {2022}}
  (\bibinfo {year} {2022})}\BibitemShut {NoStop}%
\bibitem [{\citenamefont {{Samantha Murphy Kelly}}(2023)}]{lawexam}%
  \BibitemOpen
  \bibfield  {author} {\bibinfo {author} {\bibnamefont {{Samantha Murphy
  Kelly}}},\ }\href@noop {} {\bibinfo {title} {{ChatGPT} passes exams from law
  and business schools}},\ \bibinfo {howpublished}
  {\url{https://edition.cnn.com/2023/01/26/tech/chatgpt-passes-exams/index.html}}
  (\bibinfo {year} {accessed January 2023})\BibitemShut {NoStop}%
\bibitem [{\citenamefont {{OpenAI}}(2023{\natexlab{b}})}]{gpt4}%
  \BibitemOpen
  \bibfield  {author} {\bibinfo {author} {\bibnamefont {{OpenAI}}},\
  }\href@noop {} {\bibinfo {title} {{ChatGPT}}},\ \bibinfo {howpublished}
  {\url{https://openai.com/research/gpt-4}} (\bibinfo {year} {accessed April
  2023}{\natexlab{b}})\BibitemShut {NoStop}%
\bibitem [{\citenamefont {Kortemeyer}(2023)}]{kortemeyer23ai}%
  \BibitemOpen
  \bibfield  {author} {\bibinfo {author} {\bibfnamefont {G.}~\bibnamefont
  {Kortemeyer}},\ }\bibfield  {title} {\bibinfo {title} {Could an
  artificial-intelligence agent pass an introductory physics course?},\
  }\href@noop {} {\bibfield  {journal} {\bibinfo  {journal} {Phys. Rev. Phys.
  Educ. Res.}\ }\textbf {\bibinfo {volume} {x}},\ \bibinfo {pages} {xxx}
  (\bibinfo {year} {2023})}\BibitemShut {NoStop}%
\bibitem [{\citenamefont {West}(2023)}]{west2023ai}%
  \BibitemOpen
  \bibfield  {author} {\bibinfo {author} {\bibfnamefont {C.~G.}\ \bibnamefont
  {West}},\ }\href@noop {} {\bibinfo {title} {{AI} and the {FCI}: Can {ChatGPT}
  project an understanding of introductory physics?}} (\bibinfo {year}
  {2023}),\ \Eprint {https://arxiv.org/abs/2303.01067} {arXiv:2303.01067
  [physics.ed-ph]} \BibitemShut {NoStop}%
\bibitem [{\citenamefont {Mitros}\ \emph {et~al.}(2013)\citenamefont {Mitros},
  \citenamefont {Paruchuri}, \citenamefont {Rogosic},\ and\ \citenamefont
  {Huang}}]{mitros2013}%
  \BibitemOpen
  \bibfield  {author} {\bibinfo {author} {\bibfnamefont {P.}~\bibnamefont
  {Mitros}}, \bibinfo {author} {\bibfnamefont {V.}~\bibnamefont {Paruchuri}},
  \bibinfo {author} {\bibfnamefont {J.}~\bibnamefont {Rogosic}},\ and\ \bibinfo
  {author} {\bibfnamefont {D.}~\bibnamefont {Huang}},\ }\bibfield  {title}
  {\bibinfo {title} {An integrated framework for the grading of freeform
  responses},\ }in\ \href@noop {} {\emph {\bibinfo {booktitle} {The Sixth
  Conference of MIT's Learning International Networks Consortium}}}\ (\bibinfo
  {year} {2013})\BibitemShut {NoStop}%
\bibitem [{\citenamefont {Reif}\ \emph {et~al.}(1976)\citenamefont {Reif},
  \citenamefont {Larkin},\ and\ \citenamefont {Brackett}}]{reif1976}%
  \BibitemOpen
  \bibfield  {author} {\bibinfo {author} {\bibfnamefont {F.}~\bibnamefont
  {Reif}}, \bibinfo {author} {\bibfnamefont {J.~H.}\ \bibnamefont {Larkin}},\
  and\ \bibinfo {author} {\bibfnamefont {G.~C.}\ \bibnamefont {Brackett}},\
  }\bibfield  {title} {\bibinfo {title} {Teaching general learning and
  problem-solving skills},\ }\href@noop {} {\bibfield  {journal} {\bibinfo
  {journal} {American Journal of Physics}\ }\textbf {\bibinfo {volume} {44}},\
  \bibinfo {pages} {212} (\bibinfo {year} {1976})}\BibitemShut {NoStop}%
\bibitem [{\citenamefont {Reif}(1995)}]{reif1995}%
  \BibitemOpen
  \bibfield  {author} {\bibinfo {author} {\bibfnamefont {F.}~\bibnamefont
  {Reif}},\ }\bibfield  {title} {\bibinfo {title} {Millikan lecture 1994:
  Understanding and teaching important scientific thought processes},\
  }\href@noop {} {\bibfield  {journal} {\bibinfo  {journal} {American Journal
  of Physics}\ }\textbf {\bibinfo {volume} {63}},\ \bibinfo {pages} {17}
  (\bibinfo {year} {1995})}\BibitemShut {NoStop}%
\bibitem [{\citenamefont {Hsu}\ \emph {et~al.}(2004)\citenamefont {Hsu},
  \citenamefont {Brewe}, \citenamefont {Foster},\ and\ \citenamefont
  {Harper}}]{hsu2004}%
  \BibitemOpen
  \bibfield  {author} {\bibinfo {author} {\bibfnamefont {L.}~\bibnamefont
  {Hsu}}, \bibinfo {author} {\bibfnamefont {E.}~\bibnamefont {Brewe}}, \bibinfo
  {author} {\bibfnamefont {T.~M.}\ \bibnamefont {Foster}},\ and\ \bibinfo
  {author} {\bibfnamefont {K.~A.}\ \bibnamefont {Harper}},\ }\bibfield  {title}
  {\bibinfo {title} {Resource letter rps-1: Research in problem solving},\
  }\href@noop {} {\bibfield  {journal} {\bibinfo  {journal} {American journal
  of physics}\ }\textbf {\bibinfo {volume} {72}},\ \bibinfo {pages} {1147}
  (\bibinfo {year} {2004})}\BibitemShut {NoStop}%
\bibitem [{\citenamefont {Kashy}\ \emph {et~al.}(2001)\citenamefont {Kashy},
  \citenamefont {Albertelli}, \citenamefont {Ashkenazi}, \citenamefont {Kashy},
  \citenamefont {Ng},\ and\ \citenamefont {Thoennessen}}]{kashyd01}%
  \BibitemOpen
  \bibfield  {author} {\bibinfo {author} {\bibfnamefont {D.~A.}\ \bibnamefont
  {Kashy}}, \bibinfo {author} {\bibfnamefont {G.}~\bibnamefont {Albertelli}},
  \bibinfo {author} {\bibfnamefont {G.}~\bibnamefont {Ashkenazi}}, \bibinfo
  {author} {\bibfnamefont {E.}~\bibnamefont {Kashy}}, \bibinfo {author}
  {\bibfnamefont {H.-K.}\ \bibnamefont {Ng}},\ and\ \bibinfo {author}
  {\bibfnamefont {M.}~\bibnamefont {Thoennessen}},\ }\bibfield  {title}
  {\bibinfo {title} {Individualized interactive exercises: a promising role for
  network technology},\ }in\ \href@noop {} {\emph {\bibinfo {booktitle} {Proc.
  Frontiers in Education}}},\ Vol.~\bibinfo {volume} {31}\ (\bibinfo {year}
  {2001})\ pp.\ \bibinfo {pages} {1073--1078}\BibitemShut {NoStop}%
\bibitem [{\citenamefont {Kortemeyer}\ \emph {et~al.}(2008)\citenamefont
  {Kortemeyer}, \citenamefont {Kashy}, \citenamefont {Benenson},\ and\
  \citenamefont {Bauer}}]{kortemeyer08}%
  \BibitemOpen
  \bibfield  {author} {\bibinfo {author} {\bibfnamefont {G.}~\bibnamefont
  {Kortemeyer}}, \bibinfo {author} {\bibfnamefont {E.}~\bibnamefont {Kashy}},
  \bibinfo {author} {\bibfnamefont {W.}~\bibnamefont {Benenson}},\ and\
  \bibinfo {author} {\bibfnamefont {W.}~\bibnamefont {Bauer}},\ }\bibfield
  {title} {\bibinfo {title} {Experiences using the open-source learning content
  management and assessment system {LON-CAPA} in introductory physics
  courses},\ }\href@noop {} {\bibfield  {journal} {\bibinfo  {journal} {Am. J.
  Phys}\ }\textbf {\bibinfo {volume} {76}},\ \bibinfo {pages} {438} (\bibinfo
  {year} {2008})}\BibitemShut {NoStop}%
\bibitem [{\citenamefont {Risley}(2001)}]{risley2001}%
  \BibitemOpen
  \bibfield  {author} {\bibinfo {author} {\bibfnamefont {J.}~\bibnamefont
  {Risley}},\ }\bibfield  {title} {\bibinfo {title} {Motivating students to
  learn physics using an online homework system},\ }\href@noop {} {\bibfield
  {journal} {\bibinfo  {journal} {Newsletter of the APS Forum on Education}\ ,\
  \bibinfo {pages} {3}} (\bibinfo {year} {2001})}\BibitemShut {NoStop}%
\bibitem [{\citenamefont {Stelzer}\ and\ \citenamefont
  {Gladding}(2001)}]{stelzer2001}%
  \BibitemOpen
  \bibfield  {author} {\bibinfo {author} {\bibfnamefont {T.}~\bibnamefont
  {Stelzer}}\ and\ \bibinfo {author} {\bibfnamefont {G.}~\bibnamefont
  {Gladding}},\ }\bibfield  {title} {\bibinfo {title} {The evolution of
  web-based activities in physics at illinois},\ }\href@noop {} {\bibfield
  {journal} {\bibinfo  {journal} {Newsletter of the APS Forum on Education}\ ,\
  \bibinfo {pages} {7}} (\bibinfo {year} {2001})}\BibitemShut {NoStop}%
\bibitem [{\citenamefont {Dufresne}\ \emph {et~al.}(2002)\citenamefont
  {Dufresne}, \citenamefont {Hart}, \citenamefont {Mestre},\ and\ \citenamefont
  {Rath}}]{dufresne02}%
  \BibitemOpen
  \bibfield  {author} {\bibinfo {author} {\bibfnamefont {R.~J.}\ \bibnamefont
  {Dufresne}}, \bibinfo {author} {\bibfnamefont {D.}~\bibnamefont {Hart}},
  \bibinfo {author} {\bibfnamefont {J.~P.}\ \bibnamefont {Mestre}},\ and\
  \bibinfo {author} {\bibfnamefont {K.}~\bibnamefont {Rath}},\ }\bibfield
  {title} {\bibinfo {title} {The effect of web-based homework on test
  performance in large enrollment introductory physics courses},\ }\href@noop
  {} {\bibfield  {journal} {\bibinfo  {journal} {Journal of Computers in
  Mathematics and Science Teaching}\ }\textbf {\bibinfo {volume} {21}},\
  \bibinfo {pages} {229} (\bibinfo {year} {2002})}\BibitemShut {NoStop}%
\bibitem [{\citenamefont {Fredericks}(2007)}]{fredericks2007}%
  \BibitemOpen
  \bibfield  {author} {\bibinfo {author} {\bibfnamefont {C.}~\bibnamefont
  {Fredericks}},\ }\emph {\bibinfo {title} {Patterns of Behavior in Online
  Homework for Introductory Physics}},\ \href@noop {} {Ph.D. thesis},\ \bibinfo
   {school} {University of Massachusetts} (\bibinfo {year} {2007})\BibitemShut
  {NoStop}%
\bibitem [{\citenamefont {Richards-Babb}\ \emph {et~al.}(2011)\citenamefont
  {Richards-Babb}, \citenamefont {Drelick}, \citenamefont {Henry},\ and\
  \citenamefont {Robertson-Honecker}}]{richards2011}%
  \BibitemOpen
  \bibfield  {author} {\bibinfo {author} {\bibfnamefont {M.}~\bibnamefont
  {Richards-Babb}}, \bibinfo {author} {\bibfnamefont {J.}~\bibnamefont
  {Drelick}}, \bibinfo {author} {\bibfnamefont {Z.}~\bibnamefont {Henry}},\
  and\ \bibinfo {author} {\bibfnamefont {J.}~\bibnamefont
  {Robertson-Honecker}},\ }\bibfield  {title} {\bibinfo {title} {Online
  homework, help or hindrance? what students think and how they perform},\
  }\href@noop {} {\bibfield  {journal} {\bibinfo  {journal} {Journal of College
  Science Teaching}\ }\textbf {\bibinfo {volume} {40}},\ \bibinfo {pages} {81}
  (\bibinfo {year} {2011})}\BibitemShut {NoStop}%
\bibitem [{\citenamefont {Perdian}(2013)}]{perdian2013}%
  \BibitemOpen
  \bibfield  {author} {\bibinfo {author} {\bibfnamefont {D.~C.}\ \bibnamefont
  {Perdian}},\ }\bibfield  {title} {\bibinfo {title} {Early identification of
  student performance and effort using an online homework system: A pilot
  study},\ }\href {https://doi.org/10.1007/s10956-012-9423-7} {\bibfield
  {journal} {\bibinfo  {journal} {Journal of Science Education and Technology}\
  }\textbf {\bibinfo {volume} {22}},\ \bibinfo {pages} {697} (\bibinfo {year}
  {2013})}\BibitemShut {NoStop}%
\bibitem [{\citenamefont {Docktor}\ \emph {et~al.}(2016)\citenamefont
  {Docktor}, \citenamefont {Dornfeld}, \citenamefont {Frodermann},
  \citenamefont {Heller}, \citenamefont {Hsu}, \citenamefont {Jackson},
  \citenamefont {Mason}, \citenamefont {Ryan},\ and\ \citenamefont
  {Yang}}]{docktor2016}%
  \BibitemOpen
  \bibfield  {author} {\bibinfo {author} {\bibfnamefont {J.~L.}\ \bibnamefont
  {Docktor}}, \bibinfo {author} {\bibfnamefont {J.}~\bibnamefont {Dornfeld}},
  \bibinfo {author} {\bibfnamefont {E.}~\bibnamefont {Frodermann}}, \bibinfo
  {author} {\bibfnamefont {K.}~\bibnamefont {Heller}}, \bibinfo {author}
  {\bibfnamefont {L.}~\bibnamefont {Hsu}}, \bibinfo {author} {\bibfnamefont
  {K.~A.}\ \bibnamefont {Jackson}}, \bibinfo {author} {\bibfnamefont
  {A.}~\bibnamefont {Mason}}, \bibinfo {author} {\bibfnamefont {Q.~X.}\
  \bibnamefont {Ryan}},\ and\ \bibinfo {author} {\bibfnamefont
  {J.}~\bibnamefont {Yang}},\ }\bibfield  {title} {\bibinfo {title} {Assessing
  student written problem solutions: A problem-solving rubric with application
  to introductory physics},\ }\href@noop {} {\bibfield  {journal} {\bibinfo
  {journal} {Physical review physics education research}\ }\textbf {\bibinfo
  {volume} {12}},\ \bibinfo {pages} {010130} (\bibinfo {year}
  {2016})}\BibitemShut {NoStop}%
\bibitem [{\citenamefont {Burkholder}\ \emph {et~al.}(2020)\citenamefont
  {Burkholder}, \citenamefont {Miles}, \citenamefont {Layden}, \citenamefont
  {Wang}, \citenamefont {Fritz},\ and\ \citenamefont
  {Wieman}}]{burkholder2020}%
  \BibitemOpen
  \bibfield  {author} {\bibinfo {author} {\bibfnamefont {E.}~\bibnamefont
  {Burkholder}}, \bibinfo {author} {\bibfnamefont {J.}~\bibnamefont {Miles}},
  \bibinfo {author} {\bibfnamefont {T.}~\bibnamefont {Layden}}, \bibinfo
  {author} {\bibfnamefont {K.}~\bibnamefont {Wang}}, \bibinfo {author}
  {\bibfnamefont {A.}~\bibnamefont {Fritz}},\ and\ \bibinfo {author}
  {\bibfnamefont {C.}~\bibnamefont {Wieman}},\ }\bibfield  {title} {\bibinfo
  {title} {Template for teaching and assessment of problem solving in
  introductory physics},\ }\href@noop {} {\bibfield  {journal} {\bibinfo
  {journal} {Physical Review Physics Education Research}\ }\textbf {\bibinfo
  {volume} {16}},\ \bibinfo {pages} {010123} (\bibinfo {year}
  {2020})}\BibitemShut {NoStop}%
\bibitem [{\citenamefont {Bonham}\ \emph {et~al.}(2003)\citenamefont {Bonham},
  \citenamefont {Deardorff},\ and\ \citenamefont {Beichner}}]{bonham2003}%
  \BibitemOpen
  \bibfield  {author} {\bibinfo {author} {\bibfnamefont {S.~W.}\ \bibnamefont
  {Bonham}}, \bibinfo {author} {\bibfnamefont {D.~L.}\ \bibnamefont
  {Deardorff}},\ and\ \bibinfo {author} {\bibfnamefont {R.~J.}\ \bibnamefont
  {Beichner}},\ }\bibfield  {title} {\bibinfo {title} {Comparison of student
  performance using web and paper-based homework in college-level physics},\
  }\href@noop {} {\bibfield  {journal} {\bibinfo  {journal} {Journal of
  research in science teaching}\ }\textbf {\bibinfo {volume} {40}},\ \bibinfo
  {pages} {1050} (\bibinfo {year} {2003})}\BibitemShut {NoStop}%
\bibitem [{\citenamefont {Greiffenhagen}(2014)}]{greiffenhagen2014}%
  \BibitemOpen
  \bibfield  {author} {\bibinfo {author} {\bibfnamefont {C.}~\bibnamefont
  {Greiffenhagen}},\ }\bibfield  {title} {\bibinfo {title} {The materiality of
  mathematics: Presenting mathematics at the blackboard},\ }\href@noop {}
  {\bibfield  {journal} {\bibinfo  {journal} {The British journal of
  sociology}\ }\textbf {\bibinfo {volume} {65}},\ \bibinfo {pages} {502}
  (\bibinfo {year} {2014})}\BibitemShut {NoStop}%
\bibitem [{\citenamefont {{SEB Alliance, ETH Zurich}}(2023)}]{seb}%
  \BibitemOpen
  \bibfield  {author} {\bibinfo {author} {\bibnamefont {{SEB Alliance, ETH
  Zurich}}},\ }\href@noop {} {\bibinfo {title} {Save exam browser}},\ \bibinfo
  {howpublished} {\url{https://www.safeexambrowser.org/}} (\bibinfo {year}
  {accessed April 2023})\BibitemShut {NoStop}%
\bibitem [{\citenamefont {Alharthi}\ \emph {et~al.}(2015)\citenamefont
  {Alharthi}, \citenamefont {Yahya}, \citenamefont {Walters},\ and\
  \citenamefont {Wills}}]{alharthi2015}%
  \BibitemOpen
  \bibfield  {author} {\bibinfo {author} {\bibfnamefont {A.}~\bibnamefont
  {Alharthi}}, \bibinfo {author} {\bibfnamefont {F.}~\bibnamefont {Yahya}},
  \bibinfo {author} {\bibfnamefont {R.~J.}\ \bibnamefont {Walters}},\ and\
  \bibinfo {author} {\bibfnamefont {G.~B.}\ \bibnamefont {Wills}},\ }\bibfield
  {title} {\bibinfo {title} {An overview of cloud services adoption challenges
  in higher education institutions},\ }in\ \href@noop {} {\emph {\bibinfo
  {booktitle} {Workshop on Emerging Software as a Service and Analytics}}},\
  Vol.~\bibinfo {volume} {2}\ (\bibinfo {organization} {SCITEPRESS},\ \bibinfo
  {year} {2015})\ pp.\ \bibinfo {pages} {102--109}\BibitemShut {NoStop}%
\bibitem [{\citenamefont {Kortemeyer}(2016)}]{kortemeyer2016bat}%
  \BibitemOpen
  \bibfield  {author} {\bibinfo {author} {\bibfnamefont {G.}~\bibnamefont
  {Kortemeyer}},\ }\bibfield  {title} {\bibinfo {title} {The losing battle
  against plug-and-chug},\ }\href@noop {} {\bibfield  {journal} {\bibinfo
  {journal} {The Physics Teacher}\ }\textbf {\bibinfo {volume} {54}},\ \bibinfo
  {pages} {14} (\bibinfo {year} {2016})}\BibitemShut {NoStop}%
\bibitem [{\citenamefont {{OpenAI}}(2023{\natexlab{c}})}]{chatrelease}%
  \BibitemOpen
  \bibfield  {author} {\bibinfo {author} {\bibnamefont {{OpenAI}}},\
  }\href@noop {} {\bibinfo {title} {{ChatGPT Release Notes}}},\ \bibinfo
  {howpublished}
  {\url{https://help.openai.com/en/articles/6825453-chatgpt-release-notes}}
  (\bibinfo {year} {accessed April 2023}{\natexlab{c}})\BibitemShut {NoStop}%
\bibitem [{\citenamefont {Warnes}\ \emph {et~al.}(2015)\citenamefont {Warnes},
  \citenamefont {Bolker}, \citenamefont {Bonebakker}, \citenamefont
  {Gentleman}, \citenamefont {Liaw}, \citenamefont {Lumley}, \citenamefont
  {Maechler}, \citenamefont {Magnusson}, \citenamefont {Moeller}, \citenamefont
  {Schwartz},\ and\ \citenamefont {Venables}}]{gplots}%
  \BibitemOpen
  \bibfield  {author} {\bibinfo {author} {\bibfnamefont {G.~R.}\ \bibnamefont
  {Warnes}}, \bibinfo {author} {\bibfnamefont {B.}~\bibnamefont {Bolker}},
  \bibinfo {author} {\bibfnamefont {L.}~\bibnamefont {Bonebakker}}, \bibinfo
  {author} {\bibfnamefont {R.}~\bibnamefont {Gentleman}}, \bibinfo {author}
  {\bibfnamefont {W.~H.~A.}\ \bibnamefont {Liaw}}, \bibinfo {author}
  {\bibfnamefont {T.}~\bibnamefont {Lumley}}, \bibinfo {author} {\bibfnamefont
  {M.}~\bibnamefont {Maechler}}, \bibinfo {author} {\bibfnamefont
  {A.}~\bibnamefont {Magnusson}}, \bibinfo {author} {\bibfnamefont
  {S.}~\bibnamefont {Moeller}}, \bibinfo {author} {\bibfnamefont
  {M.}~\bibnamefont {Schwartz}},\ and\ \bibinfo {author} {\bibfnamefont
  {B.}~\bibnamefont {Venables}},\ }\href
  {http://CRAN.R-project.org/package=gplots} {\emph {\bibinfo {title} {gplots:
  Various R Programming Tools for Plotting Data}}} (\bibinfo {year} {2015}),\
  \bibinfo {note} {r package version 2.16.0}\BibitemShut {NoStop}%
\bibitem [{\citenamefont {{R-Project}}(2019)}]{rproject}%
  \BibitemOpen
  \bibfield  {author} {\bibinfo {author} {\bibnamefont {{R-Project}}},\
  }\href@noop {} {\bibinfo {title} {The {R} project for statistical
  computing}},\ \bibinfo {howpublished} {\url{https://www.r-project.org}}
  (\bibinfo {year} {retrieved August 2019})\BibitemShut {NoStop}%
\bibitem [{\citenamefont {Azad}\ \emph {et~al.}(2020)\citenamefont {Azad},
  \citenamefont {Chen}, \citenamefont {Fowler}, \citenamefont {West},\ and\
  \citenamefont {Zilles}}]{azad2020}%
  \BibitemOpen
  \bibfield  {author} {\bibinfo {author} {\bibfnamefont {S.}~\bibnamefont
  {Azad}}, \bibinfo {author} {\bibfnamefont {B.}~\bibnamefont {Chen}}, \bibinfo
  {author} {\bibfnamefont {M.}~\bibnamefont {Fowler}}, \bibinfo {author}
  {\bibfnamefont {M.}~\bibnamefont {West}},\ and\ \bibinfo {author}
  {\bibfnamefont {C.}~\bibnamefont {Zilles}},\ }\bibfield  {title} {\bibinfo
  {title} {Strategies for deploying unreliable ai graders in high-transparency
  high-stakes exams},\ }in\ \href@noop {} {\emph {\bibinfo {booktitle}
  {Artificial Intelligence in Education: 21st International Conference, AIED
  2020, Ifrane, Morocco, July 6--10, 2020, Proceedings, Part I 21}}}\ (\bibinfo
  {organization} {Springer},\ \bibinfo {year} {2020})\ pp.\ \bibinfo {pages}
  {16--28}\BibitemShut {NoStop}%
\bibitem [{\citenamefont {{Wolfram Research}}(2023)}]{walpha}%
  \BibitemOpen
  \bibfield  {author} {\bibinfo {author} {\bibnamefont {{Wolfram Research}}},\
  }\href@noop {} {\bibinfo {title} {{{WolframAlpha computational
  intelligence}}}},\ \bibinfo {howpublished}
  {\url{https://www.wolframalpha.com}} (\bibinfo {year} {accessed March
  2023})\BibitemShut {NoStop}%
\bibitem [{\citenamefont {Graesser}\ \emph {et~al.}(2001)\citenamefont
  {Graesser}, \citenamefont {VanLehn}, \citenamefont {Ros{\'e}}, \citenamefont
  {Jordan},\ and\ \citenamefont {Harter}}]{graesser2001}%
  \BibitemOpen
  \bibfield  {author} {\bibinfo {author} {\bibfnamefont {A.~C.}\ \bibnamefont
  {Graesser}}, \bibinfo {author} {\bibfnamefont {K.}~\bibnamefont {VanLehn}},
  \bibinfo {author} {\bibfnamefont {C.~P.}\ \bibnamefont {Ros{\'e}}}, \bibinfo
  {author} {\bibfnamefont {P.~W.}\ \bibnamefont {Jordan}},\ and\ \bibinfo
  {author} {\bibfnamefont {D.}~\bibnamefont {Harter}},\ }\bibfield  {title}
  {\bibinfo {title} {Intelligent tutoring systems with conversational
  dialogue},\ }\href@noop {} {\bibfield  {journal} {\bibinfo  {journal} {AI
  magazine}\ }\textbf {\bibinfo {volume} {22}},\ \bibinfo {pages} {39}
  (\bibinfo {year} {2001})}\BibitemShut {NoStop}%
\bibitem [{\citenamefont {VanLehn}\ \emph {et~al.}(2005)\citenamefont
  {VanLehn}, \citenamefont {Lynch}, \citenamefont {Schulze}, \citenamefont
  {Shapiro}, \citenamefont {Shelby}, \citenamefont {Taylor}, \citenamefont
  {Treacy}, \citenamefont {Weinstein},\ and\ \citenamefont
  {Wintersgill}}]{vanlehn2005}%
  \BibitemOpen
  \bibfield  {author} {\bibinfo {author} {\bibfnamefont {K.}~\bibnamefont
  {VanLehn}}, \bibinfo {author} {\bibfnamefont {C.}~\bibnamefont {Lynch}},
  \bibinfo {author} {\bibfnamefont {K.}~\bibnamefont {Schulze}}, \bibinfo
  {author} {\bibfnamefont {J.~A.}\ \bibnamefont {Shapiro}}, \bibinfo {author}
  {\bibfnamefont {R.}~\bibnamefont {Shelby}}, \bibinfo {author} {\bibfnamefont
  {L.}~\bibnamefont {Taylor}}, \bibinfo {author} {\bibfnamefont
  {D.}~\bibnamefont {Treacy}}, \bibinfo {author} {\bibfnamefont
  {A.}~\bibnamefont {Weinstein}},\ and\ \bibinfo {author} {\bibfnamefont
  {M.}~\bibnamefont {Wintersgill}},\ }\bibfield  {title} {\bibinfo {title} {The
  andes physics tutoring system: Lessons learned},\ }\href@noop {} {\bibfield
  {journal} {\bibinfo  {journal} {International Journal of Artificial
  Intelligence in Education}\ }\textbf {\bibinfo {volume} {15}},\ \bibinfo
  {pages} {147} (\bibinfo {year} {2005})}\BibitemShut {NoStop}%
\bibitem [{\citenamefont {{Google}}(2023)}]{bard}%
  \BibitemOpen
  \bibfield  {author} {\bibinfo {author} {\bibnamefont {{Google}}},\
  }\href@noop {} {\bibinfo {title} {{Google Bard}}},\ \bibinfo {howpublished}
  {\url{https://bard.google.com/}} (\bibinfo {year} {accessed April
  2023})\BibitemShut {NoStop}%
\bibitem [{\citenamefont {{NVIDIA}}(2023)}]{nvidia}%
  \BibitemOpen
  \bibfield  {author} {\bibinfo {author} {\bibnamefont {{NVIDIA}}},\
  }\href@noop {} {\bibinfo {title} {{NVIDIA AI}}},\ \bibinfo {howpublished}
  {\url{https://www.nvidia.com/en-us/ai-data-science/}} (\bibinfo {year}
  {accessed April 2023})\BibitemShut {NoStop}%
\bibitem [{\citenamefont {{Microsoft}}(2023)}]{microsoft}%
  \BibitemOpen
  \bibfield  {author} {\bibinfo {author} {\bibnamefont {{Microsoft}}},\
  }\href@noop {} {\bibinfo {title} {{Microsoft Copilot}}},\ \bibinfo
  {howpublished} {\url{https://www.microsoft.com/en-us/ai}} (\bibinfo {year}
  {accessed April 2023})\BibitemShut {NoStop}%
\end{thebibliography}%

\end{document}